\documentstyle[11pt,aussois2003,twoside,epsf]{article}
\def\edcomment#1{\iffalse\marginpar{\raggedright\sl#1\/}\else\relax\fi}
\reversemarginpar
\begin{document}
%
%%---------------------------------------------
%%  please put the title of your talk below :
\title{Reducing and constraining the intrinsic galaxy alignment contamination to weak lensing measurements}
%%----------------------------------------------
%
%%----------------------------------------------
%% your name and affiliation :
\author{Catherine Heymans}
\affil{University of Oxford, Astrophysics, Keble Road, Oxford, OX1 3RH, UK.}
%%----------------------------------------------

%%----------------------------------------------
%% if needed, name and affiliation of your co-author :
\author{Alan Heavens}
\affil{Institute for Astronomy, University of Edinburgh, Royal Observatory,
Blackford Hill, Edinburgh, EH9 3HJ, UK.}
%%-----------------------------------------------
\label{page:first}
\begin{abstract}

We present a method for removal of the contaminating effects of
intrinsic galaxy alignments, in measurements of cosmic shear from
multi-colour weak lensing surveys. The method down-weights pairs
which are likely to be close in three dimensions, based on
spectroscopic or photometric redshifts. Results are dramatic: the
intrinsic contamination of the low-redshift Sloan photometric
redshift survey could be 80 times the lensing signal, but this
can be almost completely removed, leaving random shot noise
errors of the order of 10\%. Intrinsic galaxy alignments,
although an annoying contaminant for cosmic shear studies, are
interesting in their own right, and we therefore present a new
observational constraint for their amplitude from the
aperture mass B-mode in the Red-Sequence Cluster Survey (RCS;
Hoekstra, Yee \& Gladders 2002b). The small measured B-mode 
rules out the published intrinsic alignment models from
numerical simulations, which assume no evolution in galaxy
clustering.

\end{abstract}

\section{Introduction}

A key assumption for cosmic shear studies is that galaxy
ellipticities are randomly orientated on the sky.  It is possible,
however, that gravitational interactions during galaxy formation
could in fact produce intrinsic shape correlations between nearby
galaxies, mimicking to an extent weak lensing shear
correlations.  This effect will limit the accuracy of
cosmological parameter estimation from weak lensing studies, but
to what extent is still uncertain.  In this conference
proceedings, we review models for intrinsic galaxy alignments,
estimated from numerical simulations, and summarise an optimal
galaxy pair-weighting method, detailed in Heymans \& Heavens
(2003), which significantly reduces the contamination of weak
lensing measurements by intrinsic galaxy alignments. We propose a
simple optimised scheme for multi-colour surveys and show its
effect for the Sloan photometric survey design and the RCS survey
design.

Current analytical and numerical estimates for the amplitude and
correlation length of intrinsic galaxy alignments, although in
broad agreement on their effect on weak lensing studies, differ
in the details by an order of magnitude or more, (see contribution
by King in this volume). Intrinsic galaxy alignments are
potentially very interesting, providing information for galaxy
formation, and possibly galaxy evolution, if evolution with
redshift is observed in the intrinsic alignment signal.  It is
therefore desirable to determine the extent of intrinsic galaxy
alignments observationally. At low redshifts, where weak lensing
shear correlations are negligible, galaxy ellipticity
correlations have been observed in the SuperCOSMOS survey,
(Brown et al. 2002), and in the Tully catalogue (Pen, Lee \&
Seljak 2002). With deeper multi-colour surveys, for example
COMBO-17 (Brown et al. 2003), intrinsic galaxy alignments can be
observationally constrained by comparing ellipticity correlations
before and after their removal with 
the application of a close pair downweighting
scheme, or by using correlation function tomography (King \& Schneider
2003). An additional observational constraint that we 
focus on here arises from observed aperture mass B modes in weak
lensing measurements, which serve as a good diagnostic for the
presence of non-lensing sources within the data.

\section{Estimating intrinsic galaxy alignments from numerical simulations}

Assuming that luminous matter forms galaxies in all dark matter halos
above some minimum mass limit, it is possible to acquire a three
dimensional catalogue of galaxy shapes from a large N-body dark matter
numerical simulation.  The shape of
each luminous galaxy can be modelled
as a thin disc placed perpendicular to the angular
momentum vector of its parent dark matter halo, (Heavens, Refregier \&
Heymans 2000, HRH), or by assuming
that the galaxy ellipticity, $e$, is
equal to that of its parent dark matter halo, (Croft \& Metzler 2000;
HRH; Jing 2002).  The resulting calculated
three dimensional ellipticity correlations,  $\eta(r) = <e({\bf
  x})e^*({\bf x}+{\bf r})>$, for galaxy pairs separated
by distance $r$, show significant alignment, $\langle e e^* \rangle
\approx 0.01$,  between galaxies pairs
closer than a few tens of Mpc.  This result alone shows that in
numerical simulations, there is a
preferred angular momentum direction for nearby dark matter halos.
Projecting $\eta(r)$ into two dimensions yields an angular ellipticity
correlation function $C_I(\theta)$, which can be directly compared to
weak lensing shear correlation functions.

\begin{equation}
C_I(\theta) = \frac{\int dz_a dz_b   \phi_{z}(z_a)
\phi_{z}(z_b)  \left[1+\xi_{gg}(r_{ab})\right]
\eta(r_{ab})}{ \int dz_a dz_b   \phi_{z}(z_a) \phi_{z}(z_b)
\left[1+\xi_{gg}(r_{ab})\right]} \, ,
\end{equation}
where $\phi_{z}$ is the survey selection function, and
$\xi_{gg}(r)$ is the two-point correlation function,
which takes into account galaxy clustering.

For shallow surveys, for example the Sloan photometric survey
with median redshift $z_m \sim 0.2$, it is generally agreed that
correlations from intrinsic galaxy alignments will far exceed the
correlations expected from weak gravitational lensing.  For deeper
surveys with $z_m \sim 1.0$, assuming weak evolution in galaxy
clustering, contamination can contribute up to 10\% of the
lensing signal.

\section{An optimal method to remove the
contamination from intrinsic galaxy alignments}

If an accurate estimate of the intrinsic ellipticity correlation strength
did exist, then it would be possible to subtract it, leaving ellipticity 
correlations induced purely by lensing.  In the absence of a good
estimate for the intrinsic correlation function it is obvious that
one can reduce the contamination from intrinsic alignments by
down-weighting nearby galaxy pairs.
This is done at the expense of increasing the shot noise
contribution from the distribution of individual galaxy ellipticities
and therefore it is best to apply an optimised weighting scheme.

Given accurate galaxy redshifts, the optimal weighting for a galaxy
pair p, derived in Heymans \& Heavens (2003), is given by;
\begin{equation}
W_p = 1 - {J_p \Lambda_1\over 1+\Lambda_2}
\end{equation}
where
\begin{equation}
\Lambda_n \equiv \sum_q J_q^2,
\end{equation}
\begin{equation}
 J(r) \equiv {f I(r)\over \sigma_{\rm pair}},
\end{equation}
$\sigma_{\rm pair}$ is the shot noise error, and $f I(r)$ is the
predicted intrinsic ellipticity correlation from a galaxy pair at
separation r, multiplied by a fractional error $f$, which takes
into account the uncertainty in the intrinsic alignment model. In
practice, to guarantee the removal of any intrinsic alignment
signal, the conservative approach is to set $f =1$ and choose the
strongest plausible intrinsic alignment model for $I(r)$.

We apply this scheme to the Sloan spectroscopic survey design
comparing two intrinsic alignment models from HRH, and Jing (2002).
Figure 1 shows that this
optimal weighting scheme can largely remove the
contamination by intrinsic alignments leaving almost pure shot noise. The
shallow depth of this spectroscopic sample however,
still prevents its use for
weak lensing studies as the remaining shot noise exceeds the
expected weak lensing signal.

\begin{figure}
\plotone{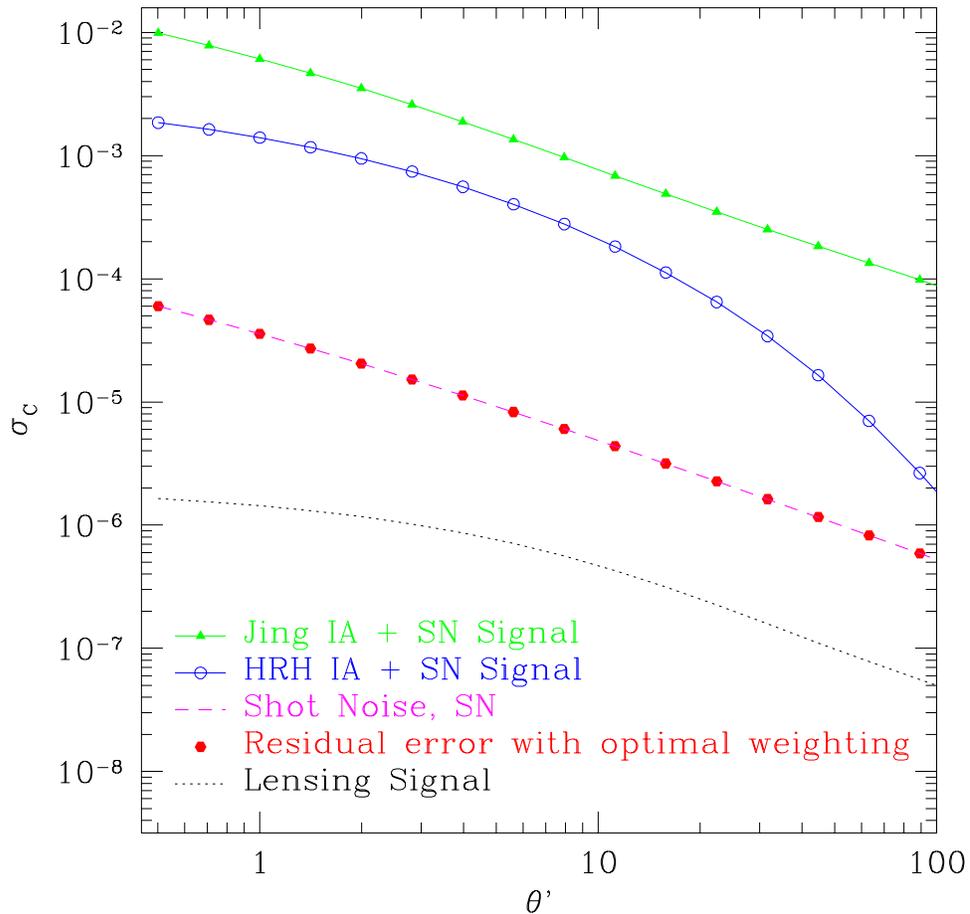}
\caption{Reduction in the error from intrinsic alignments and
shot noise for the Sloan spectroscopic sample.  The upper curves
show the error in the shear correlation function for two
intrinsic alignment models, Jing (triangles) and HRH (circles),
assuming all galaxy pairs are weighted equally.
With optimal weighting (filled),
both intrinsic alignment signals are reduced
close to shot noise (shown dashed) producing a negligible increase in
shot noise. The expected weak
gravitational lensing correlation function for a $\Lambda$CDM
model, (dotted), is still dominated by shot noise for this
shallow $z_m = 0.1$ sample.  }
\end{figure}

\section{Application to multi-colour surveys}

The optimal weighting scheme is only applicable to spectroscopic
surveys.  More often surveys have estimates of galaxy distances
from photometric redshifts, $\hat z$, with associated errors,
$\Delta_z $. These errors are much larger than the scale over
which galaxy shapes are intrinsically correlated.  We therefore
propose a semi-optimal weighting scheme dependent on estimated
redshifts such that for a pair of galaxies with estimated
redshift $\hat z_a$ and $\hat z_b$ we assign a zero weight if $ |
\hat z_a - \hat z_b| < \alpha \Delta_z $ and a weight of one
otherwise.  $\alpha$ is chosen to minimise the total error on the
shear correlation function and will depend on the angular scale,
and survey specification.  With good photometric redshifts, this
simple procedure does almost as well as the theoretical optimum
where all galaxy distances are known, and is similar to a method
simultaneously proposed by King \& Schneider (2002).

We apply this scheme to the Sloan photometric survey design, SDSS,
with $z_m = 0.2$, where the photometric redshifts are taken to be
accurate to $\Delta_z = 0.025$.  We also apply this scheme to the
RCS (Hoekstra et al. 2002a), with $z_m = 0.56$, where the
photometric redshifts are taken to be accurate to $\Delta_z =
0.3$. Figure 2 shows the correlations we expect to find for each
survey from HRH derived intrinsic alignments and the best
reduction that can be achieved with this method.  The most
startling result comes from the SDSS where, due to the wide sky
coverage and accurate photometric redshift information it is
possible to reduce intrinsic alignment systematic errors that are
80 times the lensing signal, to a random error, introduced by
shot noise, that is only 10\% of the lensing signal.  For the RCS
survey, even with fairly inaccurate photometric redshifts it is
still possible to remove a significant systematic error,
exceeding the lensing signal at angular scales $\theta < 10$
arcminutes.  This results in a 20\% random error arising from
shot noise.  Deeper surveys have also been investigated,
resulting in 10\% systematic errors being reduced to less than
1\% random errors.

\begin{figure}
\plottwo{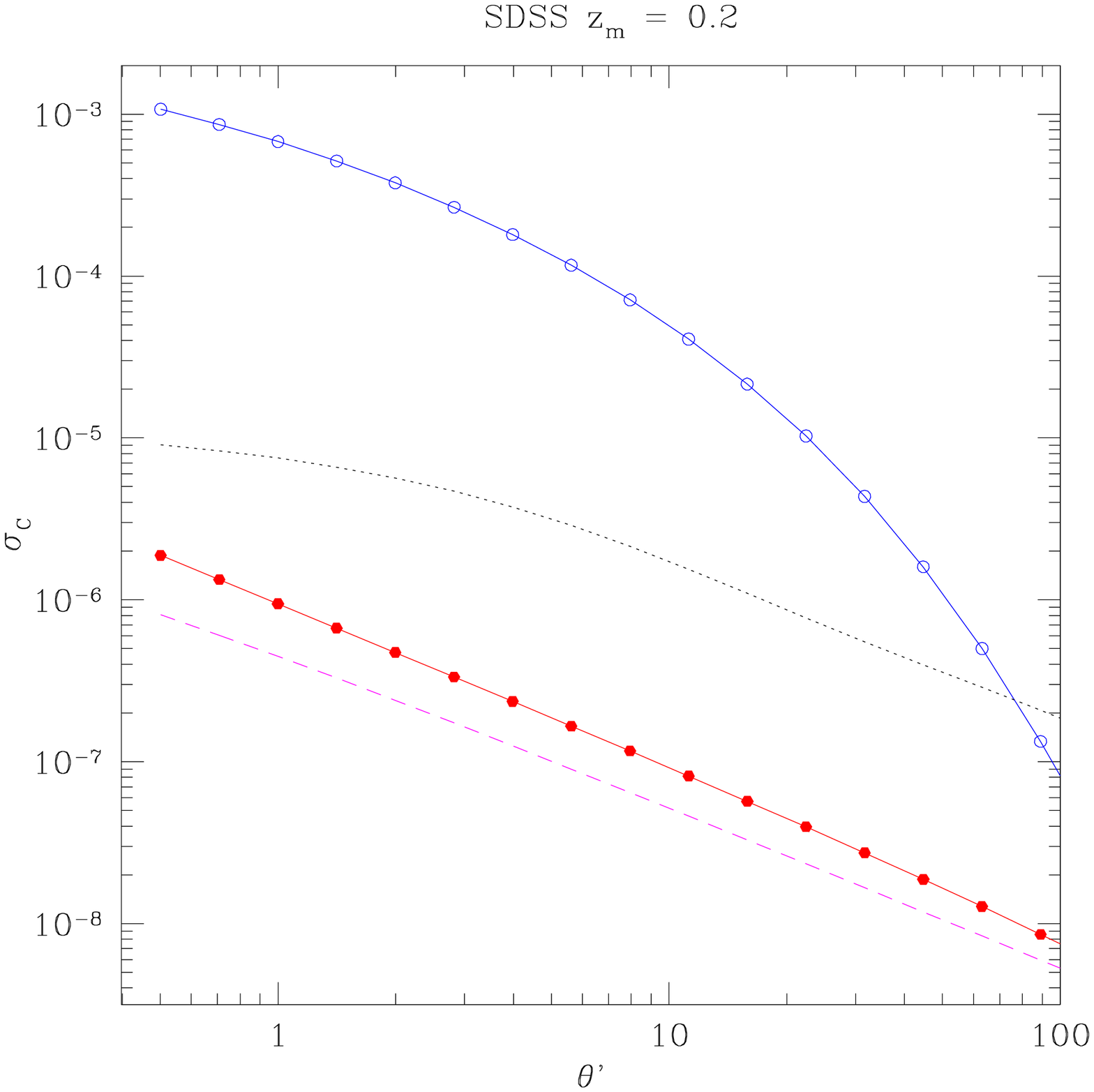}{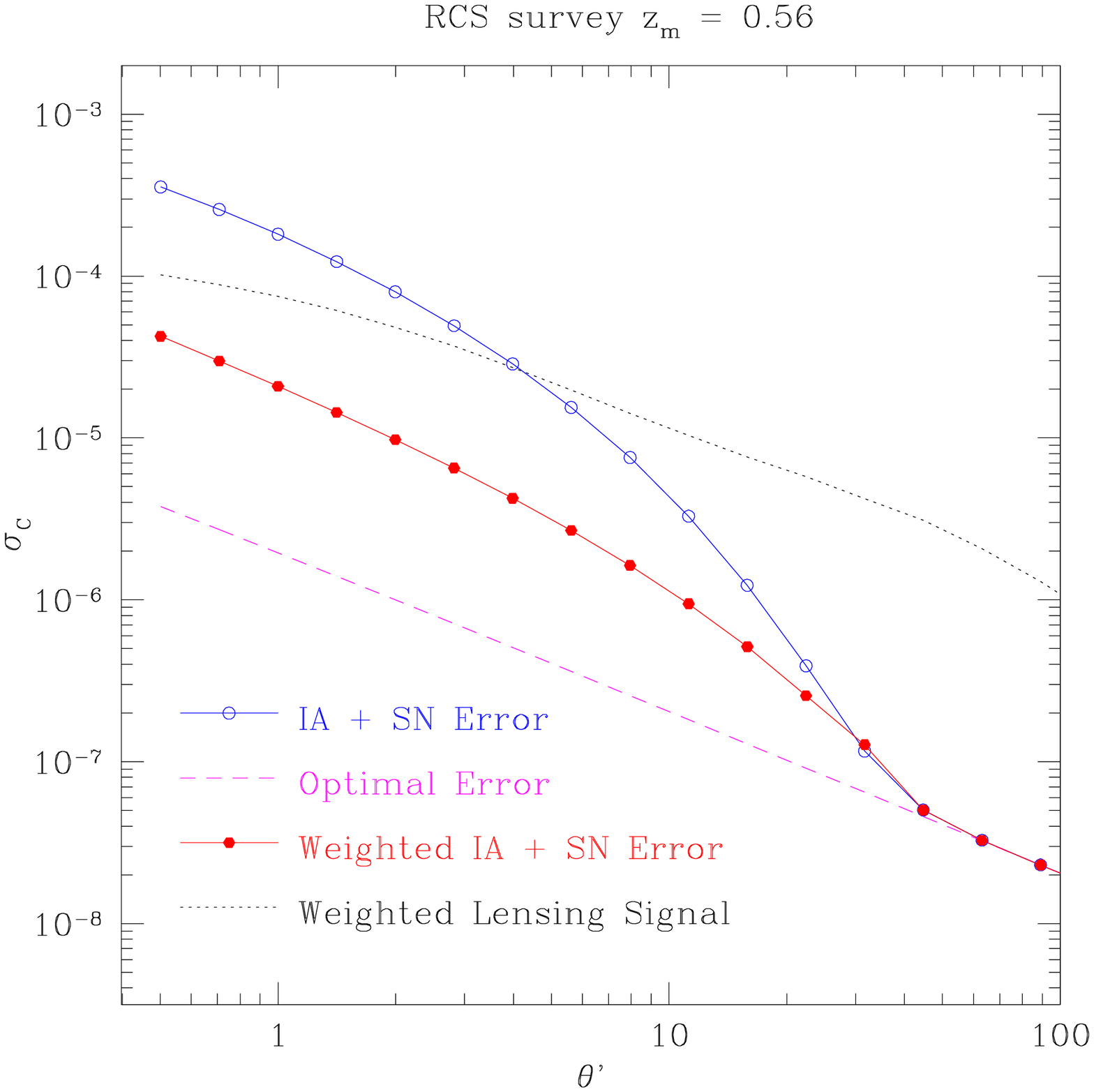}
\caption{Reduction in the error from intrinsic alignments and
shot noise for photometric SDSS and RCS.  The
semi-optimal weighting, (filled circles), has reduced the
unweighted HRH intrinsic alignment error (circles) to well below
the expected amplitude of the weighted weak lensing shear signal,
(dotted). The effect of semi-optimal weighting can be compared to
the optimally weighted error, (dashed), attainable with
spectroscopic redshifts.}
\end{figure}

\section{Aperture mass B modes. An observational constraint for intrinsic galaxy alignments}

A good diagnostic for determining the level of systematic errors
present in weak lensing measurements, is to decompose the shear
correlation signal into E and B modes. This was first proposed by
Crittenden et al. (2002), and is now a standard statistical test
for the presence of non-lensing contributions to weak lensing
measurements. Weak gravitational lensing produces curl-free
distortions (E-type), and contributes only to the B-type
distortions at small angular scales, $\theta < 1'$, due to source
redshift clustering (Schneider, Van Waerbeke \& Mellier, 2002). A
significant detection of a B-type signal in weak lensing surveys
is therefore an indication that ellipticity correlations exist
either from residual systematics within the data and/or from
intrinsic galaxy alignments which are thought to have no
preferred distortion pattern. It is therefore possible to use
observed B modes as upper limits for the B mode contribution from
intrinsic galaxy alignments, provided all data systematics
produce positive ellipticity correlations.

The decomposition for the RCS survey has been carried out
using the aperture mass statistic, $M_{ap}$, (Hoekstra et al. 2002b),
which can be directly calculated from angular
ellipticity correlation functions, (Crittenden et al. 2002;
Schneider et al. 2002).  Assuming
that intrinsic galaxy alignments have no preferred tangential or radial
alignment it is possible to calculate the expected intrinsic alignment
aperture mass B mode from the angular projection of an
intrinsic alignment model, $C_I(\theta)$.

\begin{equation}
\langle M^2_{\perp \, ({\rm IA})} \rangle = \frac{1}{2} \int_{0}^{\infty}
  \frac{d\vartheta \,
  \vartheta}{\theta^2} \left[ \xi_{+} (\vartheta) T_{+} \left(
  \frac{\vartheta}{\theta}\right) \right]
\end{equation}
where $\xi_{+}(\theta) = \frac{1}{4}C_I(\theta)$ and a useful
analytic expression for $T_{+}(x)$ is given in Schneider et al. (2002).

Figure 3 shows the intrinsic
alignment contribution from an HRH model,
to the aperture mass B mode, calculated
assuming no evolution in galaxy clustering.  Comparing this result with
the observed RCS B mode, considered as an upper limit, clearly rejects this
type of model, providing evidence that the
assumptions made for estimating
intrinsic galaxy alignment models from numerical simulations are too
simplistic, or that assuming no evolution in galaxy
clustering is incorrect.

\begin{figure}
\plotfiddle{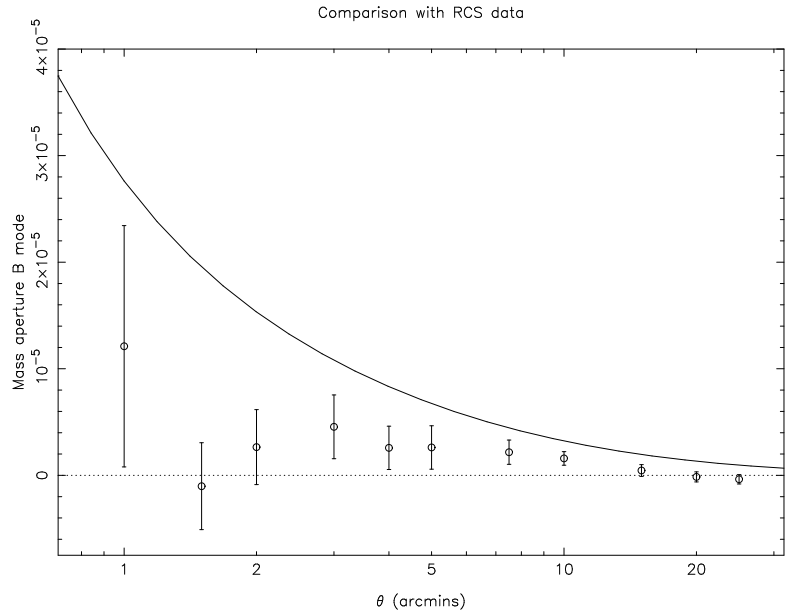}{7.5cm}{270}{42.5}{42.5}{-165}{245}
\caption{HRH intrinsic alignment model contribution to the aperture mass B mode
$\langle M_{\perp}^2 \rangle$ for the RCS survey, compared to the
  measured RCS B mode.}
\end{figure}

\section{Conclusion}

This contribution and the contribution by Lindsay King have shown
that with some redshift information it is possible to separate
galaxy ellipticity correlations induced by weak gravitational
lensing from contaminating intrinsic ellipticity correlations,
thereby removing an unknown systematic error from cosmic shear
analysis. We have presented a new observational constraint from
the aperture mass B mode statistic, prompting the need for a
re-assessment of the assumptions made when estimating intrinsic
galaxy alignments from numerical simulations. 
This constraint suggests that intrinsic alignments are a
less significant contaminant at high redshifts than was
previously predicted from numerical simulations.  Intrinsic
galaxy alignments can be directly constrained observationally at
low redshifts, for example in the Sloan survey, although we have
also shown here that this survey could now potentially be used as
a weak lensing survey with the application of an optimised close
galaxy pair down-weighting scheme.  Deeper surveys with
photometric information can also now be used to constrain
intrinsic galaxy alignments, potentially providing valuable
information for galaxy formation and evolution studies.

\section{Acknowledgements}

We thank Henk Hoekstra, Andi Burkert, Paul Allen, and Michael
Brown for useful discussions.

\label{page:last}
\end{document}